\documentclass[11pt,letterpaper]{article}
\usepackage[pass]{geometry}

\usepackage{fullpage}
\usepackage{authblk}
\usepackage[english]{babel}
\usepackage[utf8x]{inputenc}
\usepackage{amsmath}
\usepackage{amssymb}
\usepackage{graphicx}
\graphicspath{ {images/} }
\usepackage{longtable}
\usepackage{multirow}
\usepackage{hyperref}
\usepackage{subcaption}
\usepackage{comment}
\usepackage{mathtools}
\usepackage[table,xcdraw]{xcolor}
\let\oldFootnote\footnote
\newcommand\nextToken\relax

\renewcommand\footnote[1]{%
    \oldFootnote{#1}\futurelet\nextToken\isFootnote}

\newcommand\isFootnote{%
    \ifx\footnote\nextToken\textsuperscript{,}\fi}

\providecommand{\keywords}[1]{\textbf{\textit{Index terms---}} #1}

\newcommand\newsubcap[1]{\phantomcaption%
       \caption*{\thefigure(\thesubfigure)}}

\title{Align-gram : Rethinking the Skip-gram Model for Protein Sequence Analysis}
\author[1]{Nabil Ibtehaz}
\author[1]{S. M. Shakhawat Hossain Sourav}
\author[2]{Md. Shamsuzzoha Bayzid}
\author[2,*]{\\M. Sohel Rahman}

\affil[1]{Samsung R\&D Institute, Bangladesh}
\affil[2]{Department of CSE, BUET,\protect\\ ECE Building, West Palasi, Dhaka-1205, Bangladesh}
\affil[*]{Corresponding author}

\affil[ ]{}
\affil[ ]{\href{mailto:msrahman@cse.buet.ac.bd}{\nolinkurl{msrahman@cse.buet.ac.bd}}}

\begin{document}
\maketitle

\begin{abstract}

Background: The inception of next generations sequencing technologies have exponentially increased the volume of biological sequence data. Protein sequences, being quoted as the `language of life', has been analyzed for a multitude of applications and inferences. 

Motivation: Owing to the rapid development of deep learning, in recent years there have been a number of breakthroughs in the domain of Natural Language Processing. Since these methods are capable of performing different tasks when trained with a sufficient amount of data, off-the-shelf models are used to perform various biological applications. In this study, we investigated the applicability of the popular Skip-gram model for protein sequence analysis and made an attempt to incorporate some biological insights into it.

Results: We propose a novel $k$-mer embedding scheme, Align-gram, which is capable of mapping the similar $k$-mers close to each other in a vector space. Furthermore, we experiment with other sequence-based protein representations and observe that the embeddings derived from Align-gram aids modeling and training deep learning models better. Our experiments with a simple baseline LSTM model and a much complex CNN model of DeepGoPlus shows the potential of Align-gram in performing different types of deep learning applications for protein sequence analysis.

\end{abstract}


\clearpage

\section{Introduction}

With the rapid development of sequencing technology, there has been an exponential increase in the amount of biological sequence data \cite{larranaga2006machine}. This abundant amount of sequence data enables us to apply big data and deep learning-based technologies in biological sequence analysis, which is revolutionizing almost all the different aspects of life \cite{manyika2011big}. Most notably, the application of deep learning in protein sequence analysis is showing great promises \cite{larranaga2006machine,min2017deep,wen2020deep}. Previously, machine learning-based methods trained with biological features have been used in tasks like secondary structure prediction \cite{selbig1999decision}, surface residues \cite{yan2004two},  protein subcellular location \cite{huang2004prediction} etc. Deep learning-based approaches are, arguably, much superior and convenient over the traditional machine learning based bioinformatics analysis pipeline as they make the feature extraction step unnecessary and are capable of working with the raw data instead. Since the identification and extraction of such biological features are time-consuming, expensive, and require a lot of effort, deep learning-based methods are potentially more effective to keep up with the ever-increasing number of protein sequences. As a result, more and more deep learning models are being developed to perform tasks like contact prediction \cite{di2012deep}, protein secondary structure prediction \cite{netsurfp}, protein function prediction \cite{kulmanov2020deepgoplus}, protein sub-cellular localization \cite{almagro2017deeploc}, peptide-MHC binding prediction \cite{zeng2019quantification} etc.

The selection of the input representation is the first step in leveraging deep learning approaches in protein sequence analysis. The idea is to break the protein sequences into tokens (amino acids or $k$-mers) and assign a numeric value or vector to each of the tokens. This assignment can be done using several approaches. The simplest and most popular method is to apply one-hot encoding \cite{wen2020deep}, that assumes that different amino acids or $k$-mers are completely uncorrelated with each other and accordingly, treats them independently. The second-most widely used approach is to use the BLOcks SUbstitution Matrix (BLOSUM)\cite{henikoff1992amino} that encodes the amino acids using their corresponding row in the BLOSUM matrix \cite{o2018mhcflurry,jin2019attention}. This incorporates some evolutionary information of the amino acids as BLOSUM values signify which pairs of amino acids are more likely to transpose into one another during the course of evolution. This sense of relation between the amino acids may prove beneficial in certain analysis tasks. However, sometimes the simpler one-hot encoding performs better \cite{hein2020investigation}, and sometimes the combination of both yields superior results \cite{zeng2019quantification}. In addition to sequence-based approaches, evolutionary information like  Position-Specific Scoring Matrices (PSSM) \cite{jones1999protein,hanson2019improving}, physicochemical properties \cite{wang2019capsule,fu2019deepubi}, and handcrafted features \cite{abelin2019defining,yu2020dnnace} have also shown great promises. Although these latter representations may often yield surpassing results, 
this usually comes at the cost of a significant amount of analysis, studies, and experiments which are essential to compute them.    
When we consider the level of effort and contrast them with the almost on par results obtained from using simpler sequence-based representations in novel network architectures, their utility and enticement rather fade away.

Distributed representation of words has been one of the most influential and ground-breaking works in Natural Language Processing (NLP) \cite{mikolov2013efficient,mikolov2013distributed}. Earlier works, using one-hot-encoding, treated words as independent tokens. But actually, the various words are interrelated with each other, some are similar like `Perfect' and `Excellent', some are opposite like `Swift' and `Sluggish', whereas some are completely unrelated like `Computer' and `Penguin'. Thus, it is understandable that leveraging these relations between and among the words instead of treating them as completely separate entities will greatly benefit in modeling. Although the idea of generating representations for words has been investigated quite early \cite{rumelhart1986learning}, the simplistic nature of one-hot-encoding retained its acceptance, mostly because, such simple approaches proved to be robust when trained on a large amount of data \cite{mikolov2013efficient}. Nevertheless, with their revolutionary works, Mikolov et al. popularized the concept of word embedding \cite{mikolov2013efficient,mikolov2013distributed}. Now, the word2vec \cite{goldberg2014word2vec} and doc2vec \cite{lau2016empirical} models are being applied in multitude of applications. Like all other domains working with text data, bioinformatics has also embraced the potentials and possibilities of such word embedding models \cite{asgari2015continuous,kimothi2016distributed,ng2017dna2vec,yang2018learned}. In the context of bioinformatics, the concept of words is translated to $k$-mers, and using a large database of protein sequences instead of text corpus, embeddings are learned in an unsupervised manner. This type of approach has also obtained success in certain applications \cite{mazzaferro2017predicting,phloyphisut2019mhcseqnet,vielhaben2020usmpep}.

Despite the success of such approaches, in this work, we have made an attempt to investigate whether off-the-shelf NLP concepts like the popular and widely adopted skip-gram model are compatible with bioinformatics tasks like protein sequence analysis. Deep neural networks being the universal function approximator, it is always possible to model any problem with them, provided that we have a sufficient amount of data. Nevertheless, we sought out means to coalesce some biological insights with the concepts of NLP. While doing so, we identified some potential pitfalls of applying off-the-shelf Skip-gram models for protein sequence analysis and proposed a novel model, Align-gram, which is likely to be more suitable for bioinformatics applications.

\section{Methodology}
We first present a brief description of the popular Skip-gram model.
\subsection{The Skip-gram model}
The skip-gram model \cite{mikolov2013efficient} was developed to generate embeddings for words such that the syntactic and semantic relations between the words are preserved. The model constructs a vector space and maps the different words with the objective of mapping similar words close to each other. To achieve this a shallow neural network with a single hidden layer is used, that tries to predict the nearby words of a given input word. To train the model a large text corpus is used, the sentences are tokenized into words and represented by a unique id. For each word $w$ in the sentences we consider a window, i.e., we take some words ${W}$ before and after $w$, and train the model to output those words (${W}$) when $w$ is given as input. Since the words are organized in natural human sentences obeying specific grammatical and semantic rules, this procedure, in essence, allows us to approximate the meaning of the word $w$. After training the model the weights from the hidden layer are used to generate embeddings for the words.

\subsection{Motivations and Higher Level Considerations}
\label{sec:motivations}

In an earlier section, we have mentioned that off-the-shelf Skip-gram models have been applied in protein sequence analysis. However, upon diligent inspection, we point out several aspects of the Skip-gram model that might not be compatible with protein sequence analysis. First of all, in bioinformatics, we consider $k$-mers as tokens instead of words, which is the logical thing to do. However, unlike natural language where words as tokens are well defined, for $k$-mers in protein sequence it is not. Often when working with 3-mers, the common practice is to shift the sequence 2 times and break them into $k$-mers. For example, for a sequence $ADTIVAVET$, we have the 3 shifted sequences $ADTIVAVET$, $DTIVAVET$ and $TIVAVET$. We can break the 3 sequences into sets of $k$-mers as $\{ADT,IVA,VET\}$, $\{DTI,VAV\}$, $\{TIV,AVE\}$. This approach has been followed by \cite{asgari2015continuous,yang2018learned}, but we can observe that the same sequence is resulting in different contextual representations. This kind of situation is rare for natural texts as the sentences follow specific sets of rules. Furthermore, even if we consider contexts this way, we need to verify that the model manages to map the similar $k$-mers in the vector space properly, based on this contextual information about the $k$-mers is given. Although a study \cite{buchan2019inferring} investigated the semantic meaning latent in the position or context for protein domains, to the best of our knowledge there has been no study that experiments with the semantic representation obtained from $k$-mer contexts in protein sequences. We inspect the vector space generated for $k$-mers using the Skip-gram model and discover that there is hardly a correlation between $k$-mer similarity scores and the vector similarity scores of the embeddings (please refer to Section \ref{sec:correlation}). Moreover, homo-repeats \cite{lobanov2016non} like $AAAAAAAA$ exists in protein sequences, but the Skip-gram model does not consider repetitions as it is seldomly found in natural sentences.

\subsection{Align-gram}

The question that has driven us is whether the contextual information obtained from the location of $k$-mers in a protein sequence is sufficient to infer the relations between the $k$-mers. Traditionally the similarity between protein sequences or $k$-mers is determined using alignment scores \cite{peter2018encyclopedia}. During the course of evolution for point mutation, one amino acid can be transformed into another, the likelihood of which can be deduced from a substitution matrix, like BLOSUM \cite{henikoff1992amino}. Furthermore, gaps in the alignment correspond to indels, i.e., insertion or deletion mutations, which is regulated with suitable gap penalties. As a result, for decades, alignment scores are being used by biologists to measure the degree of similarity between protein sequences.

Therefore, to map the $k$-mers based on their similarity information we incorporate this biological intuition into the Skip-gram model. Instead of working with the contextual information of the surrounding $k$-mers, we compute the alignment scores of all the $k$-mer pairs. We then modify the Skip-gram model to become a regression model, by using linear activation for the output layer. Since cascaded linear layers collapse into a single linear layer, we use sigmoid activation function for the hidden layers. We represent the input $k$-mers to this network using one-hot encoding similar to Skip-gram, but unlike Skip-gram, we try to predict the alignment scores of all the $k$-mers with the input $k$-mer. We obtain a weight matrix from training the model and derive the embeddings for the $k$-mers from it, similar to Skip-gram model. 

In this way, modifying the Skip-gram model, we develop Align-gram to map the $k$-mers based on their alignment scores instead of the contextual information.

\begin{figure}[h]
     \centering
     \begin{subfigure}[b]{0.295\textwidth}
         \centering
         \includegraphics[width=\textwidth]{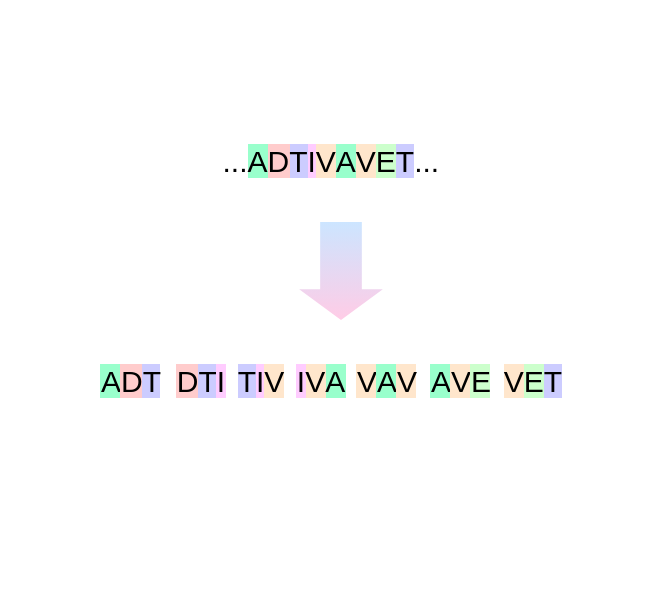}
         \caption{}
         \label{fig:break_k_mer}
     \end{subfigure}
     \hfill
     \begin{subfigure}[b]{0.295\textwidth}
         \centering
         \includegraphics[width=\textwidth]{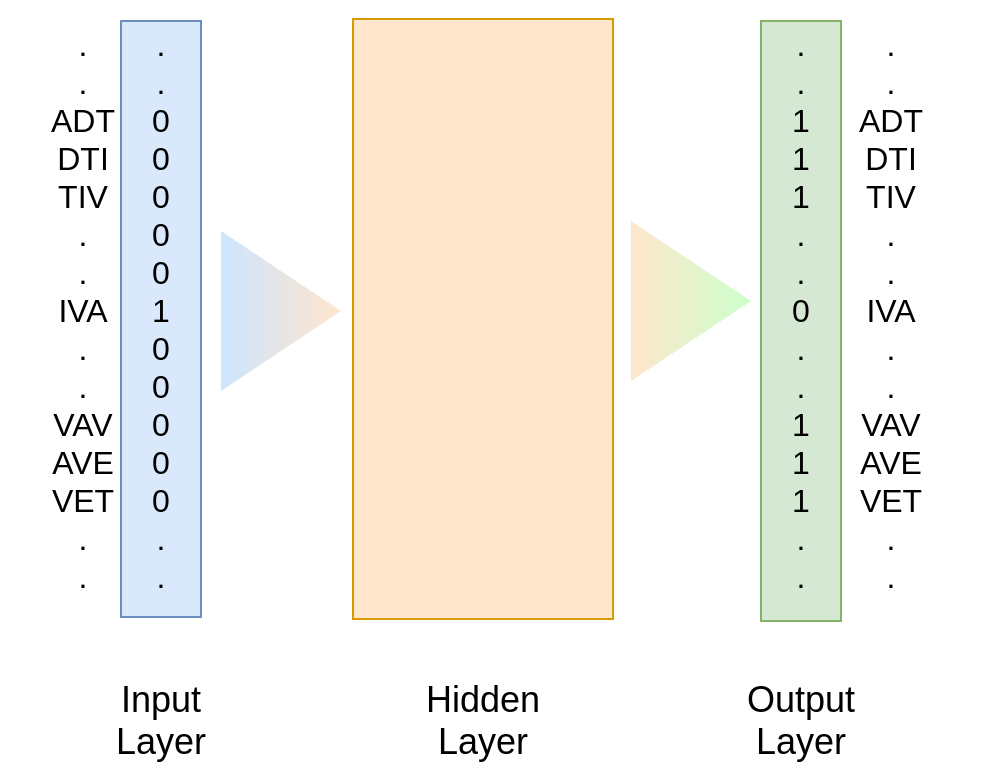}
         \caption{}
         \label{fig:skip_gram}
     \end{subfigure}
     \hfill
     \begin{subfigure}[b]{0.295\textwidth}
         \centering
         \includegraphics[width=\textwidth]{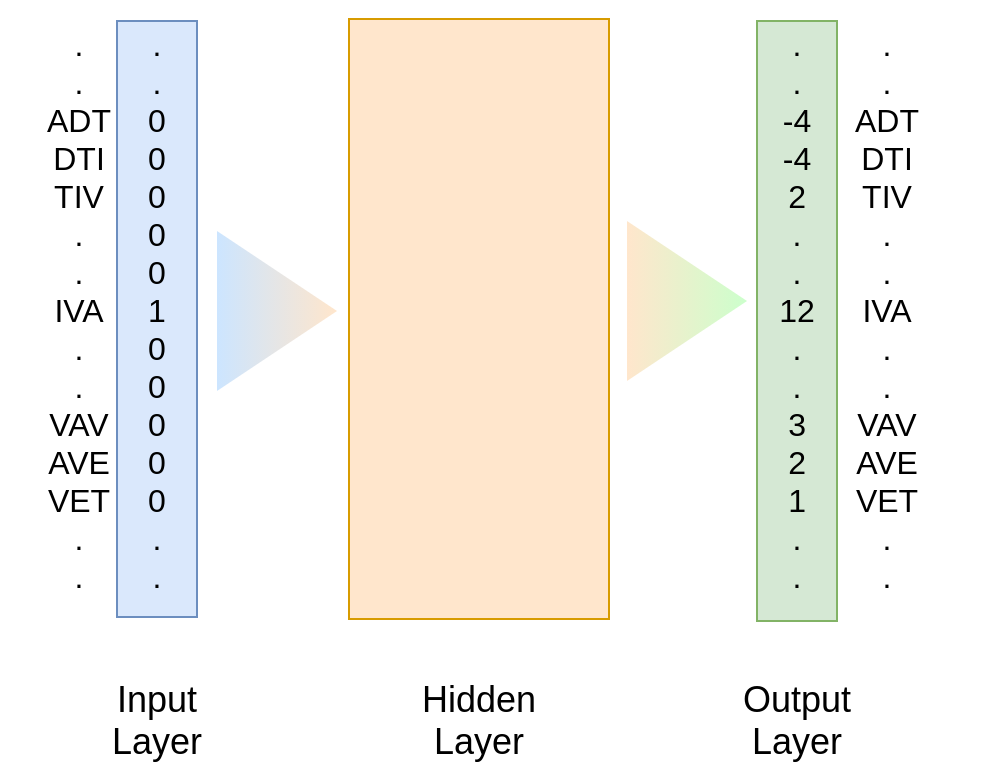}
         \caption{}
         \label{fig:align_gram}
     \end{subfigure}
        \caption{\textbf{Overview of the Process.} \ref{fig:break_k_mer} presents breaking the sequence into 3-mer. In Skip-gram model (\ref{fig:skip_gram}) the 3-mers are then used to train the model using the proximity of the 3-mers. For the proposed Align-gram model (\ref{fig:align_gram}) we compute the alignment scores between all the 3-mers and train the model to predict the alignment scores and in the process obtain the embedding matrix for 3-mers.}
        \label{fig:overview}
\end{figure}

\subsection{Implementation Details}
We computed the alignment scores between the 3-mers using the BioPython \cite{cock2009biopython} library. The values for gap open penalty, gap extend penalty were taken 11 and 1 respectively and we used the BLOSUM62 substitution matrix, which are the default options for BLASTP \cite{blast_default}.

The shallow neural network with a single hidden layer has been implemented in Keras \cite{chollet2015keras} with Tensorflow \cite{abadi2016tensorflow} backend. The hidden layer consists of 100 neurons which corresponds to the length of the embeddings. We used sigmoid and linear as the activation functions for the hidden layer and output layer respectively. The input one-hot vector was mapped to its corresponding alignment score vector as a regression problem. We minimized the mean squared error using the Adam \cite{kingma2014adam} optimizer. The model was trained for 5000 epochs, which took about 30 minutes in a Tesla P-100 GPU.

\section{Results}

\subsection{Align-gram Establishes Equivalence Between Vector Similarity and Alignment Score}
\label{sec:correlation}

The primary objective of word embeddings is to generate such a vector representation for a word so that in the vector space it is close to the vectors representing words with similar meaning. For example, we want a word like `Pizza' to be close to words like `Burger' or `Sandwich', but far from words like `Car' or `Computer' in the vector space.

\begin{figure}[h]
     \centering
     \begin{subfigure}[b]{0.45\textwidth}
         \centering
         \includegraphics[width=\textwidth]{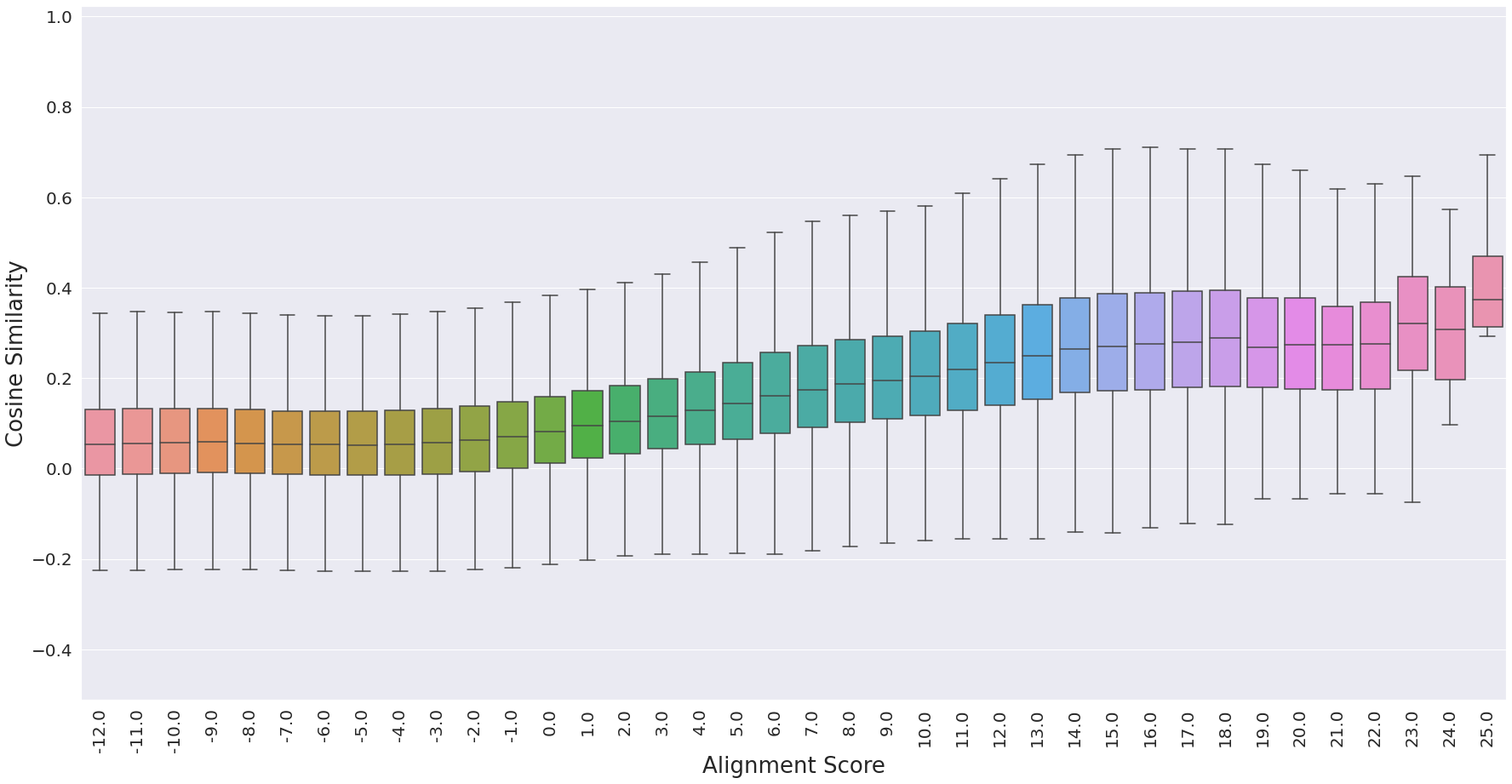}
         \caption{}
         \label{fig:cosine_sim_skipgram}
     \end{subfigure}
     \hfill
     \begin{subfigure}[b]{0.45\textwidth}
         \centering
         \includegraphics[width=\textwidth]{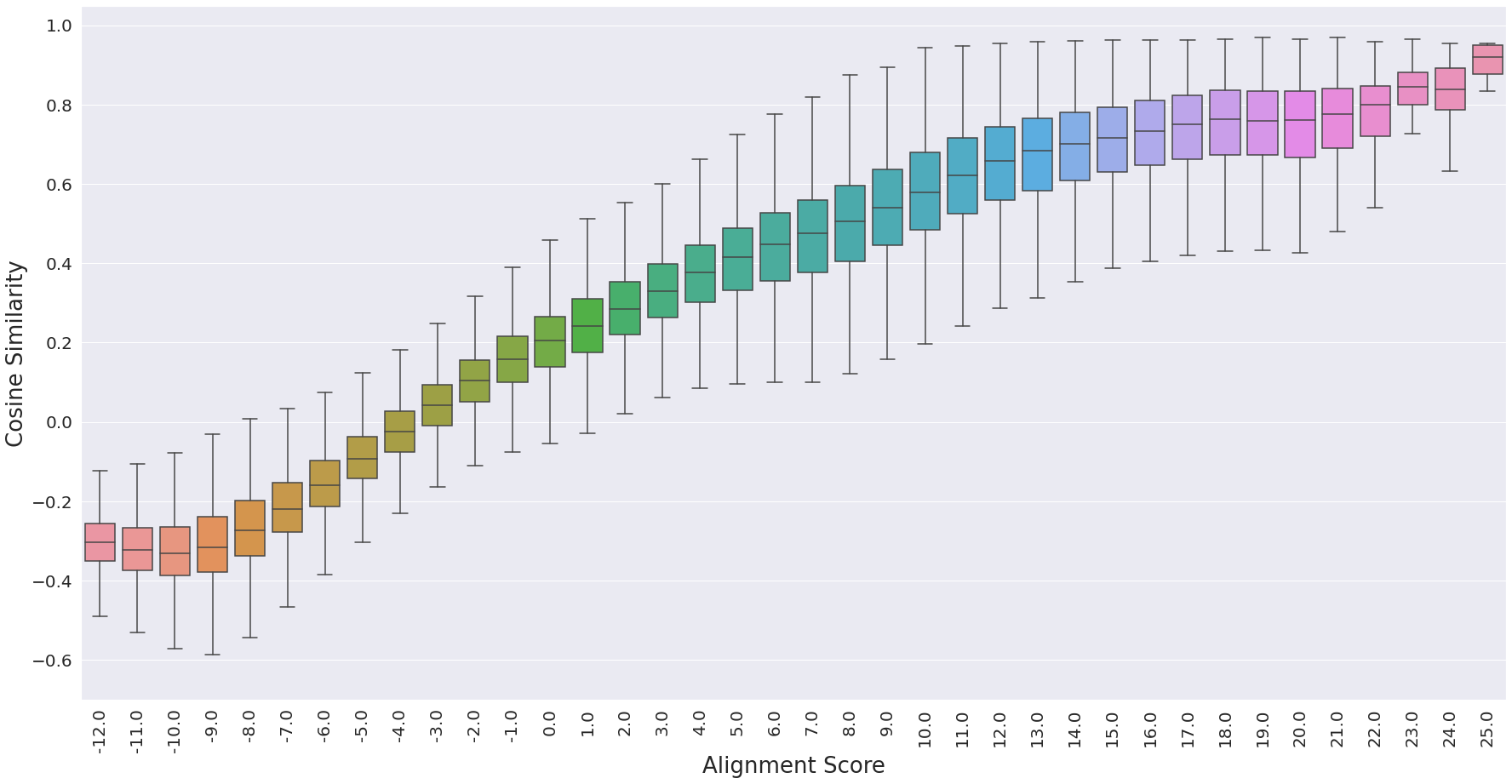}
         \caption{}
         \label{fig:cosine_sim_aligngram}
     \end{subfigure}
        \caption{\textbf{Equivalence between vector similarity and alignment score.} We present the correlation between the two using box-plots for both Skip-gram(\ref{fig:cosine_sim_skipgram}) and Align-gram(\ref{fig:cosine_sim_aligngram}) based embeddings. For Align-gram embeddings we observe a strong correlation with a Pearson correlation coefficient of 90.2\%. For better visualization the outlier of the box-plot has been omitted.}
        \label{fig:correlation}
\end{figure}

Therefore, for $k$-mer embeddings we expect the $k$-mers to be placed near similar $k$-mers. The degree of similarity between vectors and $k$-mers is computed using cosine similarity and alignment score respectively. Thus for both Skip-gram and Align-gram embeddings we compute the alignment score between every pair of $k$-mers (3-mers) and measure the cosine similarity scores of the corresponding vectors. We collected the 100-dimensional Skip-gram based embeddings for 3-mers from \cite{asgari2015continuous}. We study the correlation between cosine similarity and alignment score for the two embeddings and present them using box-plots in Fig. \ref{fig:correlation}. 

From the figure, it is apparent that the alignment scores and cosine similarity values for Align-gram embeddings are strongly correlated. On the other hand, there hardly exists a correlation for Skip-gram based embeddings, which is one particular concern we raised in Sec. \ref{sec:motivations}. The Pearson correlation coefficient for the two cases are 90.2\% and 22.6\% respectively, which quantitatively expresses the compelling correlation between vector similarity and alignment score for Align-gram embeddings. This signifies that our proposed Align-gram embeddings have managed to map similar $k$-mers close to each other.

\subsection{Align-gram Apparently Relates the Properties of Amino Acids in the Vector Space}

\begin{figure}[h]
    \centering
    \includegraphics[width=\textwidth]{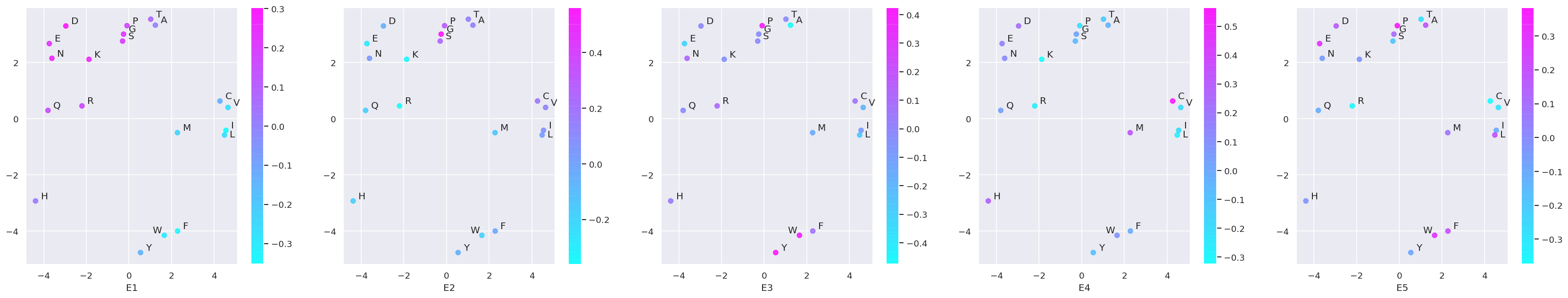}
    \caption{\textbf{Relating to the Amino Acid Properties.} We map the amino acids using Align-gram in 2 dimensional vector space and compare them with the 5 dimensional property space presented in \cite{venkatarajan2001new}. It can be observed that the eigen vector values for the nearby amino acids are close to each other.}
    \label{fig:AA_properties}
\end{figure}

In addition to maintaining the similarity between $k$-mers, we are interested to examine the capability of Align-gram to map the amino acids, preserving the physical-chemical properties. In order to do so, we train an Align-gram model with the 20 amino acids, instead of the $k$-mers. For the purpose of visualization, we compute 2-dimensional embeddings for them. Venkatarajan \cite{venkatarajan2001new} proposed quantitative descriptors for the
amino acids by summarizing 237 physical-chemical properties into a 5-dimensional property space. We collect the Eigenvectors for the amino acids from \cite{venkatarajan2001new} and compare it with our produced vector space. Fig. \ref{fig:AA_properties} presents the relation between our vector space and the summarized 5-dimensional property space. It can be observed that for most cases for nearby points the values of the eigenvectors are quite similar. For a few exceptions like $Glycine (G)$ and $Serine (S)$ we can observe that in terms of E2 and E5 values they are a bit different, but for E1, E3 and E4 they are almost identical. These minor mismatches are due to the fact that we are trying to map a 2-dimensional space into a 5-dimensional one, which itself is an approximate mapping of a 237-dimensional space. Thus, overall it is apparent that amino acid properties are conserved in our vector space.

\subsection{Align-gram Produces Better Embeddings for Protein Sequence Analysis}
\label{sec:app_psa}

In this section, we demonstrate that Align-gram based embeddings are suitable to represent protein sequences in deep learning based applications. We compare with the usual baseline sequence-based representations and perform experiments with four different datasets each considering a different type of problem.

\subsubsection{Baseline Representations}
We compare our proposed Align-gram based embeddings with the other widely used sequence based representations, namely, one-hot encoding, BLOSUM, and Skip-gram based embeddings. We represent the individual amino acids with sparse one-hot encoded vectors or corresponding rows from BLOSUM62. On the other hand, in order to use Skip-gram and Align-gram based embeddings, we tokenize the protein sequence into 3-mers and use their corresponding embedding. We have collected the 100-dimensional Skip-gram based embeddings for 3-mers from \cite{asgari2015continuous}. Hence, to keep our embeddings comparable to that of Skip-gram, we have computed 100-dimensional embeddings. Protein sequences are seldomly of uniform length, therefore to make the sequences evenly long, we pre-pad them with zeros.

\subsubsection{Tasks and Datasets}

In order to investigate the versatility of Align-gram based embeddings, we consider four different types of machine learning problems, namely, binary classification, multiclass classification, regression, and sequence-to-sequence prediction.

For the purpose of evaluating the capability of Align-gram embeddings in performing such tasks, we use four different datasets. For binary classification, we predict DNA-Binding proteins using the PDB1075 dataset \cite{liu2014idna}. For regression, multiclass classification and sequence-to-sequence prediction we use the Protein Stability Prediction \cite{rocklin2017}, Remote Homology Detection \cite{scop} and Secondary Structure Prediction \cite{pdb,casp,netsurfp} datasets from the TAPE benchmark \cite{rao2019evaluating}. We take the training split of these datasets to perform 5-fold cross-validation, and consider the PDB186 \cite{lou2014sequence} and validation split of TAPE for testing the model. All these datasets are publicly available \footnote{\url{http://server.malab.cn/Local-DPP/Datasets.html}}\footnote{\url{https://github.com/songlab-cal/tape}}.

The PDB1075 dataset consists of 1075 proteins sequences with a maximum length of 1323. The Protein Stability Prediction dataset consists of 53614 proteins whose lengths are confined to 49-50. The Remote Homology Detection dataset consists of a total of 12312 proteins with 1195 different fold labels. However, their lengths are distributed quite sporadically, the lengths cover a range of 17-1419 with a median value of 134 only. Therefore, both to mitigate the computational expense and avoid zero-padding over several magnitudes of length, we limit our analysis to sequences belonging to 75th percentile length, which equals 217. Furthermore, the dataset is highly imbalanced, hence, we only consider fold classes with at least 50 samples. This brings down the training set to 4814 sequences and similar selections were performed for the test data. The Secondary Structure Prediction dataset also suffers from a similar scattered distribution of sequence length, ranging from 20-1632 with a mean of 256, thus we similarly limit our experiments to 75th percentile length, i.e. 333.

The tasks and datasets are summarized in Table \ref{tbl:datasets}.

\begin{table}[h]
\caption{\textbf{Tasks and Datasets.} An overview of the tasks, their types and dataset information is presented here.}
\label{tbl:datasets}
\begin{tabular}{|c|c|c|c|c|}
\hline
Task & Task Type & Dataset & \# Training & \# Test \\ \hline
DNA-Binding Protein Prediction & Binary Classification & PDB 1075 & 1075 & 186 \\ \hline
Remote Homology Detection & Multiclass Classification & TAPE Benchmark & 4814 & 260 \\ \hline
Protain Stability Prediction & Regression & TAPE Benchmark & 53614 & 2512 \\ \hline
Secondary Structure Prediction & Sequence-to-Sequence & TAPE Benchmark & 6526 & 2170 \\ \hline
\end{tabular}
\end{table}
\subsubsection{Baseline Model}
\label{sec:base_model}
Since the objective of this work has been developing a representation for protein sequences, it is beyond our scope to design suitable deep network architectures for solving the different problems and properly tune their hyperparameters. Therefore, we develop a simple baseline model and experiment with the different baseline representations, using the same model. 

The baseline model consists of 3 LSTM layers each with 256 cells. 2 fully connected layers with 128 and 64 neurons respectively follow them, each with relu activation function and 20\% dropout. After these layers, we define the output layer, for regression and binary classification task the output layer consists of a single neuron with linear and sigmoid activation function respectively. For the multiclass classification problem, the output layer consists of neurons equalling the number of classes and they are activated using softmax activation function. On the other hand, for sequence-to-sequence prediction, we use time distributed fully connected layers.

\subsubsection{Experimental Results}

We train the baseline model with the different input representations for the different tasks. We evaluate the models on the test dataset, using task-specific evaluation metrics. For binary classification, we use Precision, Recall, F1, and Accuracy, for multiclass classification we use Top-1, Top-5, and Top-10 Accuracies, for regression we use Mean Squared Error (MSE), Mean Absolute Error (MAE), and Spearman's rank correlation coefficient ($\rho$), for the specific sequence-to-sequence task we compute the suitable 3 Class and 8 Class Accuracies over the entire predicted sequence. The experimental results for the different tasks are presented in Table \ref{tbl:exp}.

\begin{table}[h]
\caption{\textbf{Experimental Results.} Here, we present the results on the test data for the different tasks, using different representations. For different type of predictions suitable metrics have been used. Although the results are quite mediocre due to the simplicity of the baseline model, it can be observed that Align-gram based embeddings led to the best results (presented in bold).}
\label{tbl:exp}
\footnotesize
\begin{tabular}{|c|cccccc|}
\hline
\multicolumn{1}{|l|}{} & \multicolumn{4}{c|}{DNA-Binding-Protein Prediction} & \multicolumn{2}{c|}{Protein Secondary Structure Prediction} \\ \hline
Embedding & Precision & Recall & F1 & \multicolumn{1}{c|}{Accuracy} & 3 Class Accuracy & 8 Class Accuracy \\ \hline
One Hot & 48.11$\pm$24.33 & 60.43$\pm$31.52 & 53.24 $\pm$26.93 & \multicolumn{1}{c|}{59.89$\pm$6.30} & 50.16$\pm$1.68 & 26.80$\pm$1.73 \\
Blosum62 & 55.61$\pm$1.91 & 82.15$\pm$3.94 & 66.30$\pm$2.43 & \multicolumn{1}{c|}{58.28$\pm$2.78} & 47.37$\pm$1.14 & 25.06$\pm$1.10 \\
Skip-gram & 57.83$\pm$2.57 & 55.91$\pm$7.84 & 56.65$\pm$4.71 & \multicolumn{1}{c|}{57.63$\pm$2.75} & 46.92$\pm$3.41 & 27.58$\pm$1.57 \\
Align-gram & \textbf{61.66$\pm$4.17} & \textbf{85.59$\pm$3.23} & \textbf{71.60$\pm$3.37} & \multicolumn{1}{c|}{\textbf{65.91$\pm$5.09}} & \textbf{54.88$\pm$0.72} & \textbf{32.66$\pm$1.64} \\ \hline
\multicolumn{1}{|l|}{} & \multicolumn{3}{c|}{Remote Homology Detection} & \multicolumn{3}{c|}{Protein Stability Prediction} \\ \hline
Embedding & Top-1 Acc. & Top-5 Acc. & \multicolumn{1}{c|}{Top-10 Acc.} & MSE & MAE & $\rho$ \\ \hline
One Hot & 17.57$\pm$3.861 & 40.641$\pm$8.546 & \multicolumn{1}{c|}{59.183$\pm$6.81} & 0.17$\pm$0.003 & 0.31$\pm$0.003 & 0.78$\pm$0.005 \\
Blosum62 & 18.04$\pm$2.619 & 41.335$\pm$6.968 & \multicolumn{1}{c|}{56.706$\pm$4.303} & 0.20$\pm$0.004 & 0.35$\pm$0.005 & 0.74$\pm$0.007 \\
Skip-gram & 19.847$\pm$3.033 & 52.456$\pm$3.373 & \multicolumn{1}{c|}{70.032$\pm$2.245} & 0.20$\pm$0.009 & 0.35$\pm$0.007 & 0.73$\pm$0.012 \\
Align-gram & \textbf{24.308$\pm$2.055} & \textbf{55.332$\pm$2.056} & \multicolumn{1}{c|}{\textbf{71.702$\pm$3.774}} & \textbf{0.15$\pm$0.003} & \textbf{0.29$\pm$0.004} & \textbf{0.82$\pm$0.005} \\ \hline
\end{tabular}
\end{table}

From the table, it is evident that Align-gram based embeddings have obtained superior results in all the different tasks for all the evaluation metrics. The individual performance metric scores may appear mediocre, but this is due to the simplicity of the baseline model. For these experiments, our purpose was to demonstrate under similar settings and constraints, Align-gram embeddings are capable of producing better representations, which has been satisfied. Keeping the model simple has allowed us to perform the experiments in reduced time and we resorted to that as our objective has been to develop a novel representation, not novel architectures. Nevertheless, we show the potential of Align-gram in improving much complex deep learning models like DeepGoPlus \cite{kulmanov2020deepgoplus} in Section \ref{sec:deepgoplus}.

\subsection{Align-gram Embeddings Supplements Evolutionary Features}

Evolutionary features, although are the most capable ones for modeling biological sequences, are not always applicable for a number of reasons. For example, PSSM \cite{jones1999protein,hanson2019improving} is always changing with the discovery of new proteins \cite{hein2020investigation}, HMM state-transition probabilities \cite{netsurfp} requires sequence alignment which is time consuming and computationally expensive. As a result, language model based approaches are gaining more popularity for simplicity and convenience, despite the evolutionary features being more capable of modeling state of the art methods \cite{rao2019evaluating}.

Thus, it is expected that evolutionary features like PSSM or HMM will outperform simple sequence based representations including Align-gram embeddings. Nevertheless, we conduct experiments with the baseline model using evolutionary features. For Remote Homology Detection we use PSSM \cite{scop} and for Secondary Structure Prediction we use HMM \cite{netsurfp}. Using them as input we observe a significant improvement over the sequential representations, which is expected. However, most often these evolutionary features are coupled with one-hot-encoding \cite{netsurfp,he2018large,huang2019characterization}. Therefore, we examined coupling Align-gram embeddings with these features to see if Align-gram manages to contribute to the score. From the experiments, it was observed that including Align-gram embeddings with evolutionary features improves the result over using only evolutionary features or combining them with one-hot encoding, which is the popular practice. It may be noted here that since PSSM results in a 20-dimensional vector, we found that using 20-dimensional Align-gram embeddings were better to combine with PSSM. Align-gram offers us to use embeddings of various dimensions and more investigation is necessary to determine the optimal dimension for Align-gram embeddings.

In this way, though Align-gram could not surpass evolutionary features, still it is able to supplement them by improving the overall performance. The results are presented in Table \ref{tbl:evo}.

\begin{table}[h]
\caption{\textbf{Results Obtained from using Evolutionary Features.} For Remote Homology Detection we use PSSM and for Secondary Structure Prediction we use HMM features. It can be observed that using evolutionary features yields much better results compared to that obtained from the sequence based representations. Although Align-gram embeddings fail to outperform evolutionary features, using them together in unison yields more superior results.}
\label{tbl:evo}
\begin{tabular}{|c|c|c|c|c|}
\hline
 & \multicolumn{2}{c|}{Task : Remote Homology Detection} & \multicolumn{2}{c|}{Task : Secondary Structure Prediction} \\ \hline
Embedding & Top-1 Acc & Top-5 Acc & 3 Class Acc & 8 Class Acc \\ \hline
Evolutionary Features & 28.361 $\pm$ 1.641 & 59.154 $\pm$ 2.747 & 68.372 $\pm$ 2.028 & 44.061 $\pm$ 1.067 \\ \hline
Evo.+One Hot Enc. & 27.72 $\pm$ 1.793 & 55.644 $\pm$ 1.644 & 66.438 $\pm$ 1.557 & 44.012 $\pm$ 2.84 \\ \hline
Evo.+Skip-gram & 27.115 $\pm$ 2.571 & 61.683 $\pm$ 2.742 & 66.817 $\pm$ 2.497 & 44.175 $\pm$ 1.863 \\ \hline
Evo.+Align-gram & \textbf{29.802 $\pm$ 1.497} & \textbf{61.896 $\pm$ 2.376} & \textbf{70.572 $\pm$ 0.797} & \textbf{47.053 $\pm$ 1.042} \\ \hline
\end{tabular}
\end{table}

\subsection{Case Study : Improving DeepGoPlus using Align-gram}
\label{sec:deepgoplus}

As mentioned in Section \ref{sec:base_model}, since we primarily worked with protein sequence encoding, it was beyond our scope to design suitable deep network architectures for solving different problems. This limited us to conduct our experiments with a generic, boilerplate baseline model. Therefore, in order to demonstrate the efficacy and applicability of Align-gram in deep networks, we consider DeepGOPlus \cite{kulmanov2020deepgoplus}, one of the state-of-the-art methods for protein function prediction.

DeepGOPlus employs a novel deep convolutional neural network (CNN) architecture to extract motifs, that are likely indicators of possible protein functions. Furthermore, it combines the neural network predictions with Diamond \cite{buchfink2015fast} based sequence similarity predictions using a weighted sum model. DeepGOPlus shows competitive performance with contemporary best-performing methods. Moreover, they demonstrate astounding results in CAFA3 evaluation \cite{zhou2019cafa}, putting them as one of the three best scoring predictors for CCO and the second-best for BPO and MFO evaluations.

The authors of DeepGOPlus have made both their codes \footnote{\url{https://github.com/bio-ontology-research-group/deepgoplus}} and data \footnote{\url{http://deepgoplus.bio2vec.net/data/data-cafa.tar.gz}} publicly available. Therefore, we have collected their codes and reproduced their results using their provided data-splits, following the CAFA3 evaluations. DeepGOPlus uses one-hot-encoding to represent the protein sequences, thus, we replace the input scheme with our proposed Align-gram based embedding and rerun the experiments following precisely the same protocol. It should be noted that DeepGOPlus, being a model with over 54 million parameters, for the massive computational requirements, we ran the experiments with 20-dimensional embeddings generated using Align-gram for 3-mers. Also, the results reproduced from the official implementation of DeepGOPlus were slightly different from the results presented in the paper. This may be due to differences in software or library versions, floating precision in hardware, random seed, etc. Therefore, to make the comparisons in an even-ground, we compare the results obtained from using Align-gram embeddings with the results we reproduced. The authors also shared their computed diamond scores and thus we perform the weighted sum ensemble as well. The results are presented in Table \ref{tbl:deepgoplus}.

\begin{table}[h]
\caption{\textbf{Results Obtained from the DeepGOPlus Model.} We reproduce the original results presented in the DeepGOPlus paper. Furthermore, we rerun the experiments using Align-gram based protein sequence representation and also ensemble with Diamond scores. Using Align-gram improves the MFO and BPO metrics and for CCO a slight dip is noticed. The improvements are more noticeable for the results using the CNN model only as Diamond score seems to prevail the ensemble result a bit.}
\label{tbl:deepgoplus}
\centering
\begin{tabular}{|l|c|c|c|c|c|c|c|c|c|}
\hline
 & \multicolumn{3}{c|}{$F_{max}$} & \multicolumn{3}{c|}{$S_{min}$} & \multicolumn{3}{c|}{AUPR} \\ \cline{2-10} 
\multirow{-2}{*}{Method} & MFO & BPO & CCO & MFO & BPO & CCO & MFO & BPO & CCO \\ \hline
\multicolumn{10}{|c|}{Deep CNN Results} \\ \hline
{\color[HTML]{9B9B9B} In paper} & {\color[HTML]{9B9B9B} 0.420} & {\color[HTML]{9B9B9B} 0.378} & {\color[HTML]{9B9B9B} 0.607} & {\color[HTML]{9B9B9B} 9.711} & {\color[HTML]{9B9B9B} 24.234} & {\color[HTML]{9B9B9B} 8.153} & {\color[HTML]{9B9B9B} 0.355} & {\color[HTML]{9B9B9B} 0.323} & {\color[HTML]{9B9B9B} 0.616} \\ \hline
Reproduced & 0.405 & 0.380 & \textbf{0.598} & 9.835 & 24.120 & \textbf{8.192} & 0.337 & 0.314 & \textbf{0.593} \\ \hline
Align-gram & \textbf{0.420} & \textbf{0.395} & 0.589 & \textbf{9.688} & \textbf{23.986} & \textbf{8.192} & \textbf{0.353} & \textbf{0.335} & 0.587 \\ \hline
\multicolumn{10}{|c|}{Diamond Score Results} \\ \hline
{\color[HTML]{9B9B9B} In paper} & \multicolumn{1}{l|}{{\color[HTML]{9B9B9B} 0.509}} & \multicolumn{1}{l|}{{\color[HTML]{9B9B9B} 0.427}} & \multicolumn{1}{l|}{{\color[HTML]{9B9B9B} 0.557}} & \multicolumn{1}{l|}{{\color[HTML]{9B9B9B} 9.031}} & \multicolumn{1}{l|}{{\color[HTML]{9B9B9B} 22.860}} & \multicolumn{1}{l|}{{\color[HTML]{9B9B9B} 8.198}} & \multicolumn{1}{l|}{{\color[HTML]{9B9B9B} 0.340}} & \multicolumn{1}{l|}{{\color[HTML]{9B9B9B} 0.267}} & \multicolumn{1}{l|}{{\color[HTML]{9B9B9B} 0.335}} \\ \hline
\multicolumn{10}{|c|}{Ensemble with Diamond Score} \\ \hline
{\color[HTML]{9B9B9B} In paper} & {\color[HTML]{9B9B9B} 0.544} & {\color[HTML]{9B9B9B} 0.469} & {\color[HTML]{9B9B9B} 0.623} & {\color[HTML]{9B9B9B} 8.724} & {\color[HTML]{9B9B9B} 22.573} & {\color[HTML]{9B9B9B} 7.823} & {\color[HTML]{9B9B9B} 0.487} & {\color[HTML]{9B9B9B} 0.404} & {\color[HTML]{9B9B9B} 0.627} \\ \hline
Reproduced & 0.528 & 0.458 & \textbf{0.617} & {\color[HTML]{000000} 8.732} & \textbf{22.489} & 7.964 & {\color[HTML]{000000} 0.476} & 0.392 & \textbf{0.611} \\ \hline
Align-gram & \textbf{0.535} & \textbf{0.462} & 0.611 & \textbf{8.67} & 22.56 & \textbf{7.88} & \textbf{0.479} & \textbf{0.4} & 0.610 \\ \hline
\end{tabular}
\end{table}

From the results, it can be observed that using Align-gram based embeddings instead of one-hot-encoding improved the various metrics of DeepGOPlus. All the three metrics for MFO and BPO has been improved, and a comparatively slighter dip has been observed for CCO. The gains are more striking when we consider the predictions from the CNN model only, as in the ensemble results the Diamond scores seem to dominate a bit. 

Therefore, using Align-gram based embeddings has the potential to improve deep, sophisticated networks as well.

\section{Data and Software Availability}
Align-gram is available as free, open-source software at:
\begin{center}
\url{https://github.com/nibtehaz/align-gram}     
\end{center}
We have made our trained Align-gram embedding public in the same link, Furthermore, we have published Align-gram as a customizable interface, with sufficient documentation to aid researchers in the advancement of protein sequence analysis. 

The datasets used to develop and evaluate Align-gram are publicly available and the sources have been properly mentioned throughout the paper.

\section{Conclusion}

In this work, we started by analyzing the Skip-gram model thoroughly, with a dedicated focus revolving around the application on protein sequence analysis. We investigated and conjectured on certain scenarios and conditions which may hinder the relevance of the Skip-gram model for proteomic analysis. The significance and contribution of the Skip-gram model in the domain of Natural Language Processing (NLP) are indisputable. Nevertheless, we raise the question of whether off-the-shelf ideas and tools for broader NLP are sufficient for bioinformatics applications or not.

In one end we have the conventional biological pipelines, that computes handcrafted features, physicochemical properties, or works with evolutionary information involving PSSM or HMM profiles. All these require significant endeavors, but as they are built upon decades of knowledge compiled by the biologists and bioinformaticians, most often they yield interesting and superior results. On the other end, we have sequence based deep learning methods. These approaches are mostly based on off-the-shelf ideas from NLP domain, lacking sufficient tailoring for biological applications. Despite this, these methods have proven effective, owing to the volume of sequence data we have now, which even in unannotated form is adequate for unsupervised or self-supervised processes. Nevertheless, we made an attempt to coalesce these two contrasting opposite ideologies, to fuse the best of both worlds.

Thus, in this work, we have developed a novel word or $k$-mer embedding scheme for protein sequence analysis. We call it Align-gram, maintaining the analogies with the successful and popular Skip-gram model from NLP. Align-gram manages to correlate the embedding vector similarity of the $k$-mers with the alignment score, signifying that evolutionary and biologically similar $k$-mers are projected together. This finding is further strengthened by the conservation of amino acid properties in the learned vector space. In addition to bridging the gap between NLP model and Biological insights, Align-gram provides us with a better embedding to train deep learning models. Although Align-gram based embeddings are not on-par with computationally expensive PSSM and HMM based evolutionary features, the embeddings can supplement them well, improving the overall performance further.

The future directions of this research can be manifold. Firstly, we are interested in applying Align-gram embedding in solving a multitude of diverse problems. It would be interesting to analyze how Align-gram based embeddings affect the performance of the existing state of the art sequence based deep learning models for different tasks in protein sequence analysis. Furthermore, we are intrigued to perform similar investigations on more sophisticated NLP models like Elmo \cite{peters2018deep} and BERT \cite{devlin2018bert}. Very recently, some attempts of using Elmo and BERT models for protein sequences have been made \cite{heinzinger2019modeling,elnaggar2020prottrans}, albeit off-the-shelf NLP models without much specific modifications have been employed. Thus, it is worth exploring the possibilities of augmenting such models with biological insights and intuitions to develop more refined models, better suited for bioinformatics applications. Also, further experiments merging Align-gram embeddings with the myriad of existing hand-crafted features or evolutionary information will be beneficial for suitable input representation selection. Therefore, we believe that Align-gram based embeddings can be the proper surrogate for Skip-gram based embeddings for protein sequence analysis and will prove itself as an effective tool for bioinformaticians thereby.

\bibliographystyle{unsrt}
\bibliography{main.bib}

\end{document}